\documentclass[
superscriptaddress,
preprint,
showkeys,
amsmath,
amssymb,
aps,
pra,
longbibliography
]{revtex4-1}

\usepackage{graphicx}
\usepackage{dcolumn}
\usepackage{bm}
\usepackage[utf8]{inputenc}
\usepackage[german, english]{babel}
\usepackage{amsmath}    
\usepackage{graphicx}   
\usepackage{verbatim}   
\usepackage{color, soul}      
\usepackage{subfigure} 
\usepackage{hyperref}   
\usepackage[normalem]{ulem}
\DeclareMathOperator{\erfc}{erfc}
\begin{document}
\title{The Impact of Technologies in Political Campaigns}
\thanks{This article is published under the CC-BY-NC-ND license.}
\author{Moritz Hoferer}
\email{moritzhoferer@gmail.com}
\affiliation{ETH Zurich, Z\"urichbergstrasse 18, 8092 Zurich, Switzerland}
\author{Lucas B\"{o}ttcher}
\affiliation{ETH Zurich, Wolfgang-Pauli-Strasse 27, 8093 Zurich, Switzerland}
\author{Hans J. Herrmann}
\affiliation{PMMH. ESPCI, 7 quai St. Bernard, 75005 Paris, France}
\affiliation{Universidade Federal do Ceará, Departamento de Fisica, Fortaleza, Brazil}
\author{Hans Gersbach}
\affiliation{ETH Zurich, Z\"urichbergstrasse 18, 8092 Zurich, Switzerland}
\date{\today}
\begin{abstract} 
	\noindent 
	Recent political campaigns have demonstrated how technologies are used to boost election outcomes by microtargeting voters.
	We propose and analyze a framework which analyzes how political activists use technologies to target voters.
	Voters are represented as nodes of a network.
	Political activists reach out locally to voters and try to convince them.
	Depending on their technological advantage and budget, political activists target certain regions in the network where their activities are able to generate the largest vote-share gains.
	Analytically and numerically, we quantify vote-share gains and savings in terms of budget and number of activists from employing superior targeting technologies compared to traditional campaigns. 
	Moreover, we demonstrate that the technological precision must surpass a certain threshold in order to lead to a vote-share gain or budget advantage.
	Finally, by calibrating the technology parameters to the recent U.S.~presidential election, we show that a pure targeting technology advantage is consistent with Trump winning against Clinton.
\end{abstract}
\keywords{Campaigns; Opinion formation; Targeting; Dynamical systems}
\maketitle
\section{Introduction}
Advertising technologies have always been a key factor in political campaigns.
For instance, radio and television became relevant for political campaigns in the first half of the 20$^\mathrm{th}$ century~\cite{political_campaign_book}.
The beginning of the 21$^\mathrm{st}$ century marked another turning point for the use of technologies in elections.
With the rise of the internet, many political parties recognized the potential of cost-effective online advertisements with great reach~\cite{political_campaign_book,boettcher_petitions,Melo2018}.
In addition, online volunteer registration and online fundraising are used to increase campaign budgets and to promote a more active voter participation.
The technological development has also broadened possibilities for manipulation.
For instance, bot activism describes computer programs that create millions of fake accounts in online social networks to disseminate certain ideas and opinions to a wider audience through different sorts of online activities (e.g., through liking and (re-)tweeting)~\cite{Benigni2019Botivistm,Carley2018Science}.
More recently, we observed another transition towards data-driven campaigns which include microtargeting~\cite{kreiss18,bennett16,persily17,cadwalladr17}.
Microtargeting allows to personalize advertising and above all to identify potentially persuadable voters.
Besides the well-known examples of the use of Facebook-user-data by the campaigns of Ted Cruz and Donald J. Trump through Cambridge Analytica~\cite{Kelly2018TedCruz} and the Vote Leave campaign for Brexit through AggregateIQ~\cite{Guardian2018Brexit}, political parties are increasingly resorting to data-driven selection of potentially persuadable voters that activists can contact in door-to-door campaigns~\cite{bennett16}.
To do that, they link personal data and location data.
First, Cambridge Analytica claimed to categorize US citizens by the ``big five'' personality traits -- openness, conscientiousness, extroversion, agreeableness and neuroticism -- to provide character-specific advertising on Facebook.
Later, it corrected the statement on their web page and added that they had performed an audience segmentation, which suffices to connect local and personal data~\cite{sumpter2018outnumbered}.
There is a heated debate how much such methods benefit a campaign since hard evidence is missing.

Previous works on political opinion formation in social networks provided insights into the mechanisms that may lead to consensus, polarization, and fragmentation~\cite{Hegselmann2002Opinion,Redner2019Reality}.
Further studies suggest that media bias can significantly affect vote shares~\cite{Duggan2011Spatial,DellaVigna2007Fox}.
First strategic optimizations show how to convince voters in a discrete-time model with a continuous opinion space~\cite{Hegselmann2015Optimal}.
In this paper, we provide a framework that can help to assess the impact of technologies that target persuadable citizens (henceforth ``targeting technologies'') on political campaigns.
A better understanding of the impact of targeting technologies on political campaigns and political outcomes may not only help to explain them but also may suggest suitable policies and regulations~\cite{Gersbach2017}.
The results are compared to traditional campaigns which could not be based on the link of personal and local data.

Specifically, we propose and analyze a model that allows to include the use of targeting technologies in campaigns.
The targeting technology identifies persuadable voters who are currently in favor of the rival candidate.
These voters can be approached by campaign activists who try to convince them to support their own candidate.
This process is run on a network in which persuadable voters are represented as nodes.
Depending on their technological advantage and budget, political activists preferentially target those regions in the network where their influence generates the largest vote-share gains.

We quantify how vote-share gains and savings in terms of budget and number of activists depend on the precision of a certain targeting technology.
In particular, we characterize how these quantities depend on the ability of a given technology to identify regions with a certain minimum number of persuadable voters who are currently in favor of the competing opinion group.
Further, we also show that the precision of a given technology must surpass a certain threshold to generate a competitive advantage.
Finally, we apply our model to the polls of a recent U.S. presidential election campaign to estimate the technological advantage of Trump's campaign group.

The paper is organized as follows:
Sec.~\ref{sec:methods} introduces the network model and presents the mean-field approximation.
Further, Sec.~\ref{sec:results} analyzes various effects of the targeting technology and compares the model with empirical results. Sec.~\ref{sec:discussion} concludes our study.
\section{Methods}\label{sec:methods}
Our model is based on a network consisting of $N$ nodes which represent the persuadable part of the electorate.
Each node represents one persuadable voter connected to $k\in \left\lbrace 0,1,\dots, N-1 \right\rbrace$ other persuadable voters.
In accordance with Ref.~\cite{Bottcher2018}, we consider two campaign groups $A$ and $B$ and distinguish between political activists $A^{+}$ and $B^{+}$ and persuadable individuals $A^{0}$ and $B^{0}$.
All nodes are either of type $A^0$ or $B^0$ (i.e., $N=N_{A^0}+N_{B^0}$) and their corresponding fractions are $a^0 = N_{A^0}/N$ and $b^0 = N_{B^0}/N$.
Transitions from $B^0$ to $A^0$ ($A^0$ to $B^0$) are triggered by the persuasion attempts of activists $A^+$ ($B^+$).
There are $N_{A^+}$ ($N_{B^+}$) activists and we denote the fraction by $a^+=N_{A^+}/N$ ($b^+=N_{B^+}/N$).
More specifically, in a sufficiently small time interval $\mathrm{d}t$, with probability $ a^+ \mathrm{d}t$  ($b^+ \mathrm{d}t$), an activist $A^+$ ($B^+$) targets a certain node and its nearest-neighborhood, such that an activist reaches out to $k+1$ nodes.
In this selection, activists $A^+$ ($B^+$) then trigger transitions $B^0\rightarrow A^0$ ($A^0\rightarrow B^0$) with probability $\rho_A$ ($\rho_B$) for each voter reached out to.
We assume that the total number of activists $\widetilde{N}=N_{A^+}+N_{B^+}$ is much smaller than $N$.
Thus, the fractions $a^+$ and $b^+$ are much smaller than unity. 
In contrast to Ref.~\cite{Bottcher2018}, we consider the case in which budget expenses directly depend on the number of activists, since it captures the intensity of field operators which -- apart from volunteering -- use resources of the campaigns.
Most importantly, we introduce a variable to describe a technology that enables parties to identify regions which are particularly attractive for persuasion attempts.

Two remarks are in order.
First, while we focus on persuadable citizens and activists, it is clear that the majority of citizens is non-persuadable in presidential campaigns in the US for instance~\cite{jacobson15}.
Those citizens with fixed opinions as well as those who abstain can, of course, always be added to the model, but it does not matter for the evolution of the campaigns.  
Second, while we use the term ``persuadable voters''  to describe the possibility of opinion changes triggered by activists, most of the political economic literature analyzing campaigns use the term ``impressionable voters'' to describe citizens who may change their opinion when campaign money is spent  on them~\cite{Gersbach2011Campaigns}.

Campaigns last for a finite time which we denote by $T$.
During this period of time, sufficient budgets are necessary to pay the activist costs.
One activist costs $c_{A}$ and $c_{B}$, for campaign $A$ and $B$ respectively, per unit of time.
We define the time unit as the span during which every activist  is able to move to one node and tries to persuade each of the $k+1$ voters of the neighborhood.
Hence, the budget $B_{A}\left(t\right)$ decreases over time according to
\begin{equation}\label{eq:budget_decrease}
	\dot{B_{A}} \left( t \right)= - c_{A} \ a^+ .
\end{equation}
Furthermore, the corresponding necessary budget for a campaign of length $T$ is given by 
\begin{equation}\label{eq:budgetFunc}
	B_{A}^\mathrm{tot} \left( T \right) = 
	- \int_{0}^{T} \dot{B_{A}} \left( t \right) \mathrm{d}t =
	c_{A} a^\mathrm{+} T
\end{equation}
in the case of campaign group $A$, and $B^\mathrm{tot}_{B}\left( T \right)$ is analogous for campaign group $B$ \cite{boettcher14}.

\begin{figure}
	\centering
	\includegraphics[scale = 1.2]{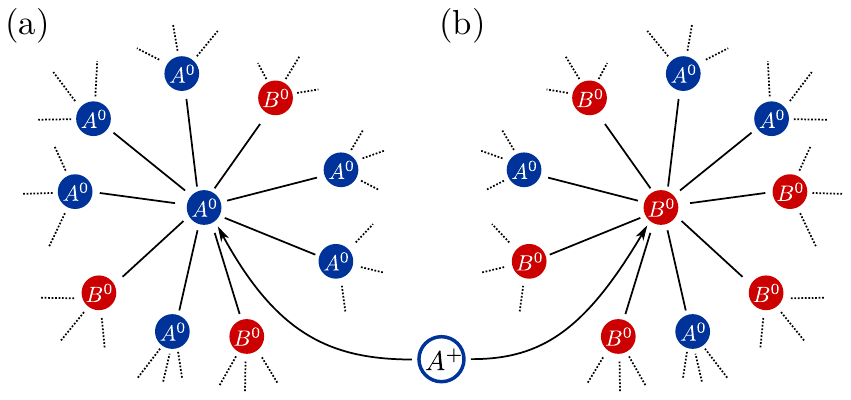}
	\caption{\textbf{Technological advantage in campaigns.}
		Both panels represent regions in a regular network of degree $k=9$. That is, an activist has the possibility to influence up to $k+1=10$ voters during one persuasion attempt. Blue nodes are of type $A^0$ and red ones of type $B^0$.
		Panel (a)~shows a vote-share situation of $A^0/B^0=7/3$ whereas  $A^0/B^0=4/6$ in panel (b).
		If campaign group $A$ ($B$) uses a technology with low precision $
		\tau \leq 0.3$ ($\tau \leq 0.4$), both central nodes are possible targets for an activist of campaign group $A$ ($B$). 
		If campaign group $A$ ($B$) uses a technology $0.3 < \tau \leq 0.6$ ($0.4 < \tau \leq 0.7$) only the central node in panel (b) ((a)) will be considered by an activist of campaign group $A$ ($B$).
		For larger technologies $\tau > 0.6$ ($\tau > 0.7$), activists $A$ ($B$) would not move to either one of the central nodes and look for nodes with higher shares of $B$ ($A$) voters.
	}
	\label{fig:technology_draft}
\end{figure}

Campaign groups have the possibility to use targeting technologies that enable them to see the network structure and to target regions where their expected vote-share gain is significantly higher~\cite{boettcher_targeted}. 
The technologies are characterized by $\tau\in \left[0,1\right]$, and we refer to $\tau$ as the precision of the technology.
The technological precision $\tau>0$ enables a campaign group to target neighborhoods with at least $\left\lceil \tau \left( k +1 \right) \right\rceil$ voters who are currently in favor of the competing opinion group, where $\left\lceil y \right\rceil$ is the ceiling function.
The case where $\tau=0$ corresponds to the situation where activists have no knowledge about the network structure and uniformly at random choose nodes.
The opposite case where $\tau = 1$ means that activists always choose the node with a number of $k+1$ persuadable individuals.
Fig.~\ref{fig:technology_draft} gives an illustrative example of the effect of a technology with precision $\tau$.

We next examine the evolution of ${a}^0 \left( t \right)$ and ${b}^0 \left( t \right)$, assuming that the two campaigns employ targeting technologies $\tau_A$ and $\tau_B$, respectively.
We focus on the mean-field rate equations of $\dot{a^0}\left(t\right)$ and $\dot{b^0}\left(t\right)$, and thus assume a perfectly mixed population in the thermodynamic limit.
Mean-field approximations are an important tool to obtain analytical results for a wide range of dynamical systems \cite{keeling-rohani2008,marro05}.
It is noteworthy that many real-world social networks are well described by mean-field approximations \cite{gleeson12}.

Activists target a certain node and its $k$ neighbors, and then try to convince these $k+1$ nodes during one persuasion attempt.
The analytical treatment of general degree distributions is described in Refs.~\cite{boettcher162,boettcher171}.
Due to the fact that $b^0 = 1 - a^0$ and $\dot{b}^0 = - \dot{a}^0$, the dynamics of $b^0$ is determined by the time evolution of $a^0$ and vice versa:

\begin{align}
	\begin{split}
		\dot{a^0} \left( t \right)
		=\quad &
		\underbrace{
		\rho_{A} a^+ \frac{ \sum_{j=\lceil \tau_{A} \left( k + 1 \right) \rceil}^{k+1} j \binom{k+1}{j} {b^0 \left( t \right)}^j {a^0 \left( t \right)}^{k+1-j}}{\sum_{j=\lceil \tau_{A} \left(k+1\right)\rceil}^{k+1} \binom{k+1}{j} {b^0 \left( t \right)}^j {a^0 \left( t \right)}^{k+1-j}}}_{
		\text{gain: } f_{B^0 \rightarrow A^0}\left( t \right) } \\
		-&
		\underbrace{ \rho_{B} b^+
		\frac{ \sum_{j=\lceil \tau_{B} \left(k+1\right) \rceil}^{k} j \binom{k+1}{j} {a^0 \left( t \right)}^j {b^0 \left( t \right)}^{k+1-j}}{\sum_{j=\lceil \tau_{B} \left(k+1\right) \rceil}^{k+1} \binom{k+1}{j} {a^0 \left( t \right)}^j {b^0 \left( t \right)}^{k+1-j}}
		}_{
		\text{loss: }f_{A^0 \rightarrow B^0}\left( t \right)}.
	\end{split}
\label{eq:voterFlow}
\end{align}
The first term describes the gain (loss) of votes and the second the loss (gain) of votes for party $A$ ($B$).
Both terms are built in the same manner.
The term $\rho_{A}$ ($\rho_{B}$) is the probability that a voter changes the opinion as result of an activist persuasion attempt.
Moreover, $a^{+}$ ($b^{+}$) is the share of party $A$ ($B$) activists in the population.
The fraction with the two sums represents the number of voters with the opposite opinion that an activist expects to meet.
According to the lowest index $\left\lceil\tau_{A} \left(k + 1\right)\right\rceil$  ($\left\lceil\tau_{B} \left(k + 1\right)\right\rceil$) in the sums, the technology constrains the set of nodes with the neighborhood the activists will visit.
This implies that an activist $A^{+}$ ($B^{+}$) will only visit neighborhoods with at least $\left\lceil\tau_{A} \left(k + 1\right)\right\rceil$ ($\left\lceil\tau_{B} \left(k + 1\right)\right\rceil$) nodes in the state $B^{0}$ ($A^{0}$).

For our subsequent analysis, we consider the situation in which both parties have a sufficiently large budget such that activists can be active in persuading voters throughout the entire campaign.
In App.~\ref{app:uniqueness}, we illustrate that the dynamics defined by Eq.~\eqref{eq:voterFlow} together with Eq.~\eqref{eq:budget_decrease} exhibits a unique steady state $a^0_{\mathrm{st}}=1-b^0_\mathrm{st}$.
Since the stationary state is globally stable, the model predicts that for a given set of parameters after infinitely long campaigns ($T\rightarrow\infty$) for any initial vote-shares $a^{0}\left( 0 \right)$ and $b^{0}\left( 0 \right)$, the two campaign groups reach vote-shares $a_\mathrm{st}^{0}$ and $b_\mathrm{st}^0$ on the election day.

We note that for limited budgets, Heaviside functions depending on the budgets $B_A \left( t \right)$ and $B_B \left( t \right)$ have to be added in Eq.~\eqref{eq:voterFlow} in front of both the gain and loss terms, respectively.
Thus, transitions $B^0 \rightarrow A^0$ and $A^0 \rightarrow B^0$ triggered by activists would only occur at times $t<T$ if there is enough budget $B_A \left( t \right)>0$ and $B_B \left(t\right) >0$ available.
\section{Results}
\label{sec:results}
The benefit of a technological advantage in a campaign is shown first conceptually and second for the U.S. presidential election in 2016.
For the first part, we focus on actual vote-share gains which result from effective targeting technologies.
We quantify the savings of an implemented technology in terms of budget and number of political activists. 
Particularly, we demonstrate that the precision of a technology must surpass a threshold in order to lead to a competitive advantage.
In order to assess the impact of a targeting technology, we assume that campaign group $A$ uses a targeting technology with precision $\tau_{A}=\tau>0$, whereas campaign group $B$ does not have access to such a technology (i.e., $\tau_{B}=0$).
We assume constant costs $c_{A}$ and $c_{B}$ and shares of activists $a^+$ and $b^+$ to guarantee analytical tractability.
In the second part, we move closer to a real world application by assessing the technological advantage of campaigns in the recent U.S.~presidential election of 2016.
\subsection{Vote-share gains due to a technological advantage}\label{sec:res_voteshare}
\begin{figure*}
	\centering
	\begin{minipage}{0.49\textwidth}
		\centering
		\includegraphics[scale=.9]{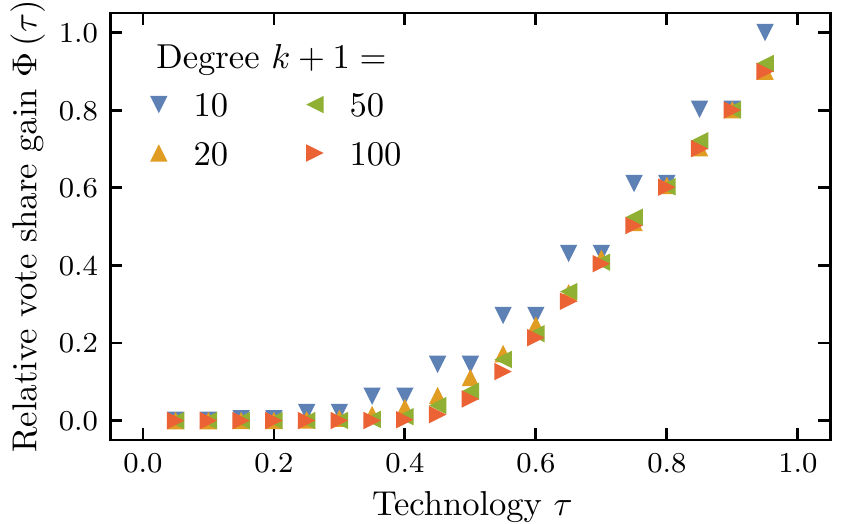}
	\end{minipage}
	\hfill
	\begin{minipage}{0.49\textwidth}
		\centering
		\includegraphics[scale=.9]{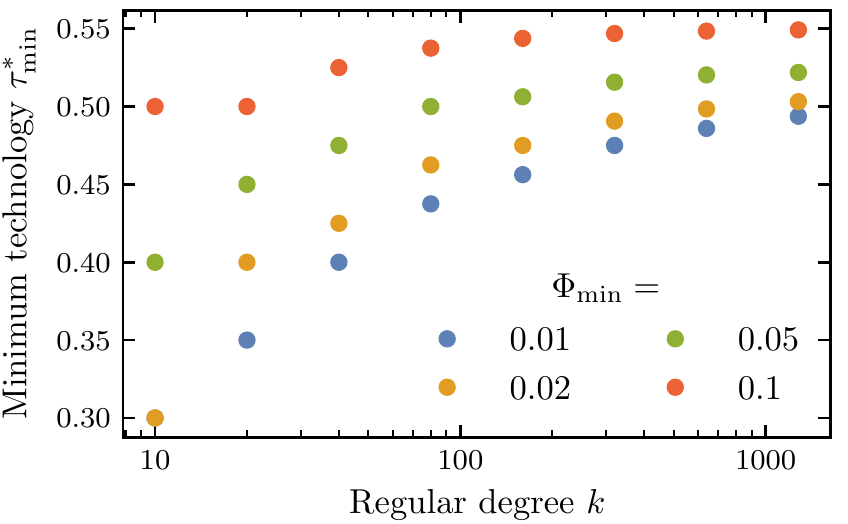}
	\end{minipage}
	  \caption{
	  	\textbf{Technology-induced vote-share gains.}
	  	The left panel shows the order parameter $\Phi(\tau)=\left[a^0_{\mathrm{st}}(\tau)-a^0_{\mathrm{st}}\left(0\right)\right]/a^0_\mathrm{st}\left( 0 \right)$ as a function of $\tau$ for $a^+=b^+$ and $\rho_{A}=\rho_{B}$.
	  	As $\tau$ exceeds a threshold value $\Phi_\mathrm{min}$, it leads to a competitive advantage for the respective campaign group. In the right panel, we show $\tau^\ast_\mathrm{min}=\min_{\tau\in[0,1]}\{\tau|\Phi(\tau)\geq \Phi_\mathrm{min}\}$ as a function of the degree $k$ for different $\Phi_\mathrm{min}$. For large degrees, we find that $\tau^\ast_\mathrm{min}\rightarrow 0.5$ as $\Phi_\mathrm{min}\rightarrow 0$.
	  	In the vicinity of $\tau^\ast$, the relative vote-share gain scales as $\Phi\sim \left( \tau - \tau^\ast \right)$.
  	 } 
	 \label{fig:tau_influence}
\end{figure*}

In order to focus on the effect of vote-share gains due to targeting technologies, we now consider the situation where all parameters of both campaign groups coincide except for $\tau$.
For that, we set $\rho = \rho_A = \rho_B$, $c_A = c_B$, and $a^+ = b^+$ as constants.
We focus on plausible ranges for these parameters:
Typically, plausible values of $\rho$ do not exceed $5 \%$, whereas the activist fractions $a^+$ and $b^+$ are smaller than $10 \%$ (see e.g.\ Ref.~\cite{huffpost}).
These values represent the fact that activists are a small share of the electorate and that the probability to persuade an persuadable voter is very low.
The campaign budgets are assumed to be large enough not to be depleted within the duration $T$ of the campaign.
In Fig.~\ref{fig:tau_influence} (left), we quantify vote-share gains according to 
\begin{equation}
	\Phi(\tau)=
	\left[a^0_{\mathrm{st}}(\tau)-a^0_{\mathrm{st}}\left( 0 \right)\right] / a^0_\mathrm{st}\left( 0 \right).
	\label{eq:phi_tau}
\end{equation}
We find that a large value of $\tau$ leads to a substantial vote-share gain since both campaigns would obtain $50 \%$ of the votes for equal targeting technologies.
If the precision $\tau$ raises to unity, the relative vote-share gain increases to one which means that the campaign group $A$ owns all vote-shares $a^0_\mathrm{st}=1$.

In reality, it might be impossible to observe the limit $\tau = 1$. This is due to the fact that with increasing network degree $k$ and decreasing electorate size $N$, there is only a vanishing probability of finding at least one node with persuadable neighbors that vote for the competing party.
However in our analysis, we focus on the transition characteristics for smaller values of $\tau$. Interestingly, for small values of $\tau$, the advantage is negligible compared to the situation where $\tau=0$.
For values of $\tau \gtrsim 0.5$, the advantage is not negligible anymore.
This finding implies that there exists a threshold that $\tau$ must surpass to have a significant impact on the dynamics.

We analytically describe this threshold effect and elaborate on the underlying mechanism in the subsequent section and in App.~\ref{app:caseDifferentiation}.
The intuition for the existence of tipping points is: A targeting technology only has an effect if it targets more voters than a random voter-selection strategy. That is, it has an effect if $\tau \geq b^0_\mathrm{st}$.
Interestingly, the transition appears to smear out for small degrees and to become more sharp for increasing $k$.

Based on the results presented in App.~\ref{app:caseDifferentiation}, the relative vote-share gain scales as $\Phi\sim \left( \tau - \tau^\ast \right)$ in the vicinity of $\tau^\ast$.
We study this effect more systematically in Fig.~\ref{fig:tau_influence} (right).
The minimum technology $\tau^\ast_\mathrm{min}$ is defined as the value of $\tau$ at which $\Phi(\tau)$ is larger than $\Phi_\mathrm{min}$. That is, $\tau^\ast_\mathrm{min}=\min_{\tau\in[0,1]}\{\tau|\Phi(\tau)\geq \Phi_\mathrm{min}\}$ where $\Phi_\mathrm{min}$ is a threshold we can choose.
We observe that $\tau^\ast_\mathrm{min}\rightarrow 0.5$ for large degrees $k$ as $\Phi_\mathrm{min}$ approaches zero.
This characterizes the expected transition indicating that $\tau^\ast_\mathrm{min}=0.5$ separates two phases in the limit of large degrees for $a_\mathrm{st}^{0} \left( 0 \right) = 0.5$.
\subsection{Relating budgets, activists and technology}\label{sec:res_budgets}
Technologies that allow campaign groups to effectively target voters lead to larger corresponding vote-shares if their technological precision is above a certain tipping point.
In contrast, a more effective technology should lead to a greater activist- and budget-saving potential.
Next, we establish a relation between the share of activists, campaign budgets and the technological precision.
According to Eq.~\eqref{eq:budgetFunc}, the campaign budgets only depend on the costs and shares of activists.

It is possible to express the relative budget $B_{A}^\mathrm{tot}/B_{B}^\mathrm{tot}$ and the relative activist fraction $a^+/b^+$ as function of the stationary fractions $a_{\mathrm{st}}^0$ and $b_{\mathrm{st}}^0$ and the precision $\tau$.
According to Eq.~\eqref{eq:budgetFunc}, the relative budget is given by
\begin{equation}\label{eq:relativeBudget}
\frac{B_A^\mathrm{tot} \left(T\right)}{B_B^\mathrm{tot}\left(T\right)}=
\frac{\int_0^T c_A a^+ \mathrm{d} t}{\int_0^T c_B b^+ \mathrm{d} t}=
\frac{c_A a^+ }{c_B b^+},
\end{equation}
where we assumed constant costs and shares of activists for the second equality.
In the stationary regime of Eq.~\eqref{eq:voterFlow} which is obtained by setting $\dot{a^0}\left( t \right) \overset{!}{=} 0$, it is possible to express the relative budget $B_{A}^\mathrm{tot}/B_{B}^\mathrm{tot}$, as defined in Eq.~\eqref{eq:relativeBudget} and the relative activist fraction $a^+/b^+$ in terms of $a^0_{\mathrm{st}}$ and $b^0_{\mathrm{st}}$:
\begin{equation}\label{eq:relativeBudgetStep2}
\frac{B_A^\mathrm{tot} \left(T\right)}{B_B^\mathrm{tot} \left( T \right)}=
\frac{c_A a^+}{c_B b^+} =
\frac{c_A \rho_B}{c_B \rho_A} 
\frac{	
	\frac{ \sum_{j=0}^{k+1} j \binom{k+1}{j} {a^0_\mathrm{st}}^j {b^0_\mathrm{st}}^{k+1-j}}{\sum_{j=0}^{k+1} \binom{k+1}{j} {a^0_\mathrm{st}}^j {b^0_\mathrm{st}}^{k+1-j}}
	}{
	\frac{\sum_{j=\lceil \tau_A \left(k +1 \right) \rceil}^{k+1} j \binom{k+1}{j} {b^0_\mathrm{st}}^j {a^0_\mathrm{st}}^{k+1-j}}{\sum_{j=\lceil \tau_A \left(k+1\right) \rceil}^{k+1} \binom{k+1}{j} {b^0_\mathrm{st}}^j {a^0_\mathrm{st}}^{k+1-j}}
	}=
	\frac{c_A \rho_B}{c_B \rho_A}\frac{\left(k+1\right) a^0_{\mathrm{st}}}{\frac{ \sum_{j=\lceil \tau_A \left(k+1\right) \rceil}^{k+1} j \binom{k+1}{j} {b^0_\mathrm{st}}^j {a^0_\mathrm{st}}^{k+1-j}}{\sum_{j=\lceil \tau_A \left(k+1\right) \rceil}^{k+1} \binom{k+1}{j} {b^0_\mathrm{st}}^j {a^0_\mathrm{st}}^{k+1-j}}}.
\end{equation}

In App.~\ref{app:caseDifferentiation}, we derive the second equality of Eq.~\eqref{eq:relativeBudgetStep2}.
In the next step, we develop an approximation for a comparably large number of connections $k$ with respect to reasonable and realistic parameters.
Typical values of the degree $k$ are in the range between 10 to 50 \cite{privatecomm}.
This not only yields further analytical insights, but may also capture actual persuasion efforts of activists through internet such as online social networks which typically display a large number of connections.
In particular, we apply a Gaussian approximation to the denominator of Eq.~\eqref{eq:relativeBudgetStep2}.

We present the detailed derivation in App.~\ref{app:caseDifferentiation} and find that the relative budget exhibits the following dependence on the technological precision $\tau$ of campaign group $A$:
\begin{equation}
	\frac{B_A^\mathrm{tot} \left( T \right)}{B_B^\mathrm{tot} \left( T \right)}=
	\frac{c_A a^+}{c_B b^+} =
	\frac{c_A \rho_B}{c_B \rho_A} \cdot
	\begin{cases}
		\frac{a^0_\mathrm{st}}{b^0_\mathrm{st}} &\text{if } \tau < b^0_\mathrm{st} = 1 - a^0_\mathrm{st}\\
		\frac{a^0_\mathrm{st}}{\tau} &\text{if } \tau \geq b^0_\mathrm{st} = 1 - a^0_\mathrm{st}\\
	\end{cases}~.
	\label{eq:reducedBudgetAnalyticalSolution}
\end{equation}
We already discussed the occurrence of a threshold effect in Sec.~\ref{sec:res_voteshare}, where $\tau$ needed to surpass a certain value in order to lead to a noticeable vote-share gain.
This effect is also contained in Eq.~\eqref{eq:reducedBudgetAnalyticalSolution}. 
We conclude that for large degrees $k$, a technological precision $\tau$ must perform better than a random strategy which, on average, targets the stationary fraction $b^0_\mathrm{st}$.
For small degrees, however, the threshold might be smaller, as shown in Fig.~\ref{fig:tau_influence}.

\begin{figure*}
	\centering
	\begin{minipage}{0.49\textwidth}
		\centering
		\includegraphics[scale=.9]{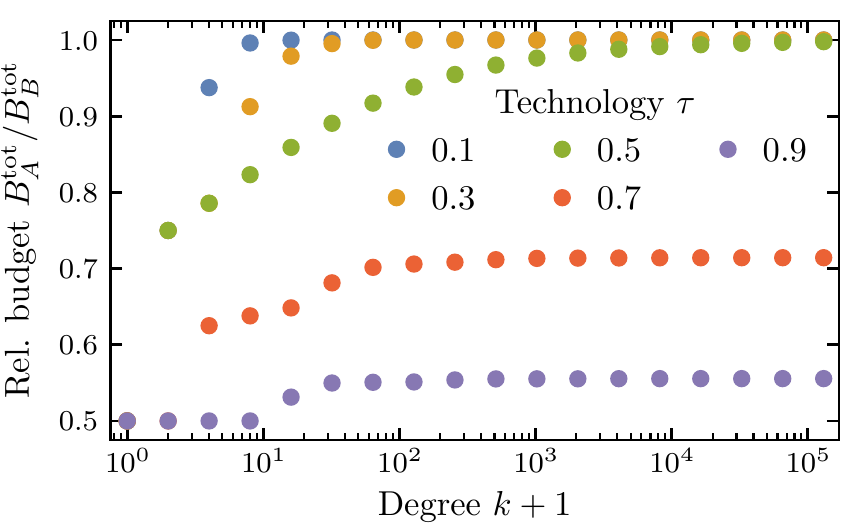}
	\end{minipage}
	\hfill
	\begin{minipage}{0.49\textwidth}
		\centering
		\includegraphics[scale=.9]{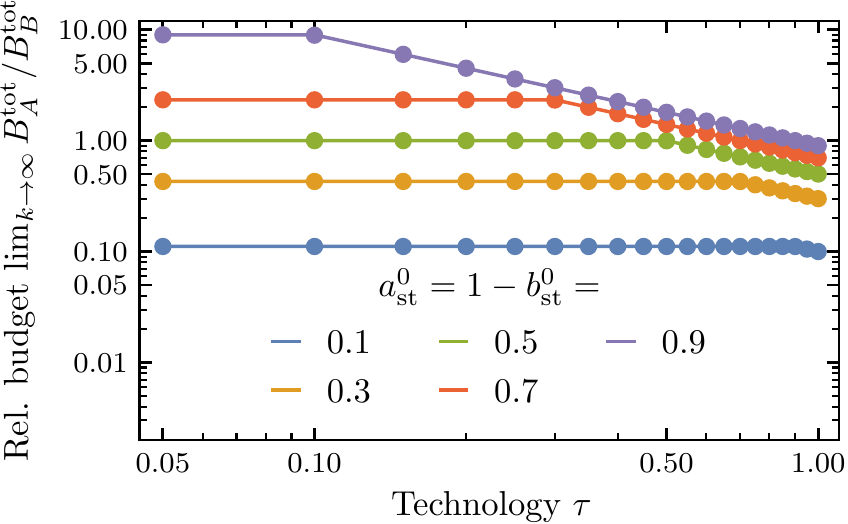}
	\end{minipage}
	\caption{
		\textbf{Relative budgets and technology.}
		In the left panel, we illustrate the relative budget $B_A^\mathrm{tot}/B_B^\mathrm{tot} \sim a^+/b^+$ for a steady state situation with $a^0_\mathrm{st} = b^0_\mathrm{st}=0.5$ and different values of $\tau$ which campaign group $A$ needs to achieve the same vote-share as campaign group $B$.
		The numerical solution is based on Eq.~\eqref{eq:relativeBudgetStep2}.
		The right panel shows the numerical solution of Eq.~\eqref{eq:relativeBudgetStep2} for $k + 1=10^5$ (lines) on logarithmic scales and the analytic solution given by Eq.~\eqref{eq:reducedBudgetAnalyticalSolution} (markers) as a function of $\tau$ for different steady state vote-shares and $\rho_A=\rho_B$ and $c_A=c_B$. 
	}
	\label{fig:reducedBudget}
\end{figure*}

In Fig.~\ref{fig:reducedBudget} (left), we depict the relative budget $B_A^\mathrm{tot}/B_B^\mathrm{tot}$ as a function of the degree $k$ for different values of $\tau$, equal steady state vote-shares $a^0_\mathrm{st} = b^0_\mathrm{st}=0.5$ and equal activist costs and persuasion probabilities for both parties.
In agreement with the arguments presented in App.~\ref{app:caseDifferentiation}, we find that the relative budget approaches a constant as the degree becomes large, such that the binomial sum of Eq.~\eqref{eq:relativeBudgetStep2} is well-described by a Gaussian approximation.
Similar to the transition in Fig.~\ref{fig:tau_influence} (left), we also find that the relative budget decreases significantly for values of $\tau\gtrsim 0.5$.
This effect is illustrated on logarithmic scales by the green curve of Fig.~\ref{fig:reducedBudget} (right).
In addition, the dependence of the relative budget on $\tau$ for four other steady state vote-share scenarios is shown in Fig.~\ref{fig:reducedBudget} (right).
For all different steady state vote-shares, the relative budget is constant at 
$a^0_\mathrm{st}/b^0_\mathrm{st}$ up to $\tau = b^0_\mathrm{st}$, and then decreases with $1/\tau$.
In agreement with Eq.~\eqref{eq:reducedBudgetAnalyticalSolution}, the relative budget increases with $a^0_\mathrm{st}$ and decreases with $\tau$ if $\tau\geq b^0_\mathrm{st}$.
\subsection{Parameter estimation for the U.S.~presidential election of 2016}
\begin{figure}
	\centering
	\includegraphics[scale=.9]{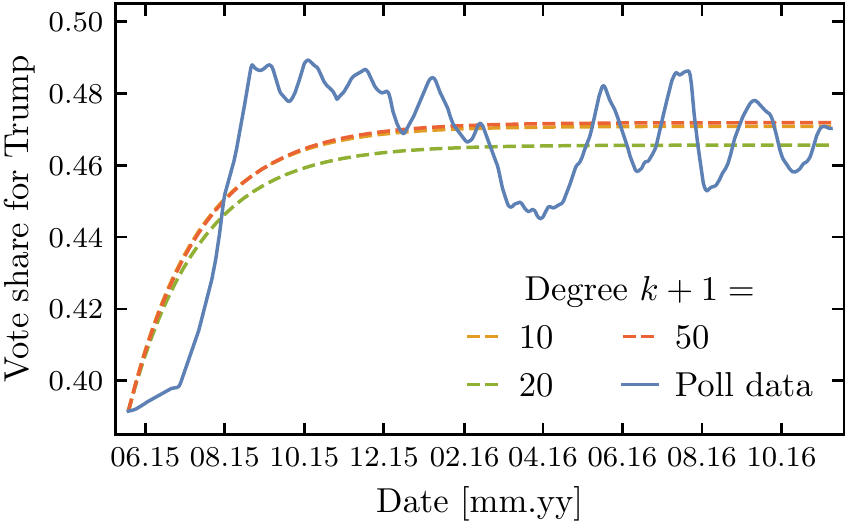}
	\caption{
		\textbf{Parameter estimation for the U.S.~presidential election of 2016.}
		The vote-share for Trump, based on poll data of the U.S.~presidential election of 2016 and the corresponding least-square fitted parameters of Tab.~\ref{tab:parameterEstimation}.
		The poll data has been taken from HuffPost Pollster \cite{pollchart}.
		The parameter estimation only refers to Trump and Clinton.
	}
	\label{fig:parameterEstimation}
\end{figure}

In this part, we apply our model to gauge the targeting technology that Trump's campaign group might have possessed to win against the Clinton campaign.
For that, we could take the election result, but it is unclear whether the result represents the stationary state.
Hence, it is better to take the dynamics of opinion formation into account.
Performing a parameter estimation for campaign dynamics is possible because of the high availability of polling data~\cite{pollchart}.
In Fig.~\ref{fig:parameterEstimation}, the solid blue line shows the dynamics of the vote-share for Trump during the one and a half year long campaign. 
For the first 6 months, the curve shows an equilibration process and after it fluctuates around the later result.
In the following calibration, we take the whole opinion dynamics of the campaign into account to estimate both the activist activity for both parties and targeting technology of the Trump campaign.
Of course, this exercise also works for the shorter period after it became clear that Clinton and Trump were the candidates of both parties.
Furthermore, since political campaigns involve a variety of influence channels beyond the activists and targeting technologies we examined in this paper, such a calibration has to be executed in an appropriate and cautious way

We proceed in several steps:
First, we use the actual data on budgets and activists that both presidential campaigns had at their disposal and note that Trump's campaign had nearly half the budget and number of activists of the Clinton campaign ($a^{+}=0.04$ [Trump], $b^{+}=0.07$ [Clinton])~\cite{insidegov}.
Additionally, we introduce a time-scale parameter $\tilde{t}$ that indicates the average time in days that passes between two persuasion attempts of one activist.
We rescale the shares of activists as $\tilde{a}^{+}=\tilde{t}^{-1} a^{+}$ and $\tilde{b}^{+}=\tilde{t}^{-1} b^{+}$, respectively.
Second, we identify the precision of the technology of the Clinton campaign by $\tau = 0$, matching traditional campaigns that could not target small groups of persuadable voters through the link of personality and location data~\cite{jacobson15}.
The results are robust to the choice of $\tau$ of the Clinton campaign in $\left[0, a_\mathrm{st}^{0}\right]$ as we have theoretically discussed in Sec.~\ref{sec:res_budgets}.
Further, we elaborate this on the data of the U.S. presidential election of 2016 at the end of this section.
Finally, we estimate the technology advantage of the Trump campaign. That is, we estimate the parameter $\tau$ that is consistent with the election result.
For this purpose, we assume that $\rho_{A}=\rho_{B}=0.05$ which puts all burden on technological edge to explain the vote-share evolution and is in line with the evidence that persuasion probabilities are very small~\cite{jacobson15}.
Allowing for a larger persuasion probability for the Trump campaign would, of course, lower the corresponding technological edge~\cite{Bottcher2018}.

To estimate the technology advantage we could use the stationary values and equate them to the election result.
It is more convenient to use the 
poll data averaged over the available polls as this allows additionally to compare the entire dynamics during the campaign.
We keep in mind that these data turned out to be biased and that most of the activist dynamics occurred in battleground states.
Our assumption is that this bias remained stable over time, such that changes in average polls reflect changes in support.
Based on the poll data of the entire campaign, we examine whether the vote-share evolution is consistent with a targeting technology advantage for the Trump campaign.

Fig.~\ref{fig:parameterEstimation} shows the poll data of Trump vs.~Clinton and fitting curves to three different degrees $k$.
The poll data is normalized by the vote-shares of Trump and Clinton.
The curves are fitted according to the numerical solution of Eq.~\eqref{eq:voterFlow} and the first data point of the time series as initial condition.

\begin{table}
	\centering
	\begin{tabular}{cccc}
		\hline
		$k + 1 = $ & $\tau\in$ &  $\tilde{t} \approx$ &  $\left\langle\Delta^2\right\rangle=$ \\ 
		\hline
		10 & $\left(0.7,0.8\right]$  & 2 & $2.47 \cdot 10^{-4}$\\ 
		20 & $\left(0.75,0.8\right]$ &  4 & $2.68 \cdot 10^{-4}$ \\ 
		50 & $\left(0.8,0.82\right]$ & 10 & $2.47\cdot10^{-4}$\\
		100  & $\left(0.81,0.82\right]$ & 19 & $2.46\cdot10^{-4}$\\
		\hline
	\end{tabular}
	\caption{
		\textbf{Parameter estimation.}
		The least-square fitted parameters $\tau$ and $\tilde{t}$ for different degrees $k$, $\rho_{A}=\rho_{B}=0.05$ and $c_A=c_B$.
		Additionally, the last column shows the error of the fit.
		The technological precision $\tau$ is given as an interval of length $1/k$, since all values of the interval, in combination with the ceiling function in Eq.~\eqref{eq:gaussApproxeval1}, yield the same result.
		The time scale $\tilde{t} \in \mathbb{N}$ indicates after how many days an activist has to travel to a new node to make a persuasion attempt.
	} 
	\label{tab:parameterEstimation}
\end{table}

Tab.~\ref{tab:parameterEstimation} presents the resulting parameters according to least squares.
The technological precision is given in intervals including all parameters that are consistent with the ceiling function.
The upper bound is the relevant value and the size of the interval scales with the degree as $k^{-1}$.
For increasing values of $k$, we observe an increase in technological precision that best fits the data.
The table illustrates that the vote-shares are consistent with the existence of a targeting technology advantage or in other words, despite the disadvantage in terms of budgets and activists, the targeting technology advantage as represented in Tab.~\ref{tab:parameterEstimation} is sufficient to yield the observed vote-share evolution and ultimately allows the Trump campaign to win the election.

The resulting estimate presented in Tab.~\ref{tab:parameterEstimation} indicates a large value of technological precision $\tau\approx 0.8$.
According to the arguments developed in Sec. \ref{sec:res_voteshare} and \ref{sec:res_budgets}, this value has to be interpreted relative to random targeting, since the competing Clinton campaign group targets a fraction of $a^0_\mathrm{st}\approx  48.9 \%$  on average.
Thus, a value of $\tau=0.8$ would correspond to a $31.1 \%$ larger technological precision of the Trump campaign in comparison to the one by Clinton.
So despite having a significant disadvantage in terms of the number of activists and the amount of budget, a pure technology advantage is sufficient to reproduce Trump's victory.

We have applied a variety of robustness checks.
First, the results are robust to changes of $\rho_{A}$ ($=\rho_{B}$) in the range of small values $\left[0.02, 0.1\right]$.
This is valid as the persuasion probability only affects the equilibration  time and the time-scale $\tilde{t}$ in the parameter estimation, respectively.
Second, since we cannot assign a precise value of $\tau$ for the Clinton campaign---since even in traditional ones some limited targeting may be  possible---we have performed the same exercise for $\tau$ values in $\left[0, a^0_\mathrm{st}\right]$  of the Clinton campaign.
From Sec.~\ref{sec:res_budgets}, we already know that the result should remain the same.
Indeed, the impact on the precision of the technology $\tau$ of the Trump in the simulation campaign are very small.
For example for $k+1=100$ and the technology of Clinton $\tau \leq 0.37$, the interval of Trump's technology remains in the range of $\left(0.81, 0.82\right]$.
\subsection{Cross-checking and Validation}\label{sec:crosschecking}
In conclusion, we discuss the main result established above and we provide an argument why even a lower technological advantage may have been sufficient if we consider the county structure.
First, adding non-persuadable voters would not affect our results.
Assuming that non-persuadable voters are distributed equally among the two campaign groups~\cite{Gentzkow2016Polarization,PewResearchCenter2018}, the result of $\tau$ does barely change, as the resulting steady state would be the same ($a^{0}_\mathrm{st} \approx b^{0}_\mathrm{st} \approx 0.5$).
The same dynamics can be achieved by appropriate re-scaling of the degree~$k$ or the time scale parameter~$\tilde{t}$.
Second, as the probability to find a node having a neighborhood with a higher vote-share for Clinton than $0.8$ is small in the mean-field approximation, the value $\tau$ of the precision of the technology seems to be too high.

For that reason, we consider the vote-share distribution of the election outcome on the county level to obtain an individual mean-field model for each county.
Every county obeys the same dynamics as described in Sec.~\ref{sec:methods}, and gets weighted by the total number of cast votes in the respective county.
Next, we analyze whether the value of $\tau$ is really necessary to achieve the observed vote-share gain when we use the empirical distribution and structure of vote-shares in each county instead of an averaged distribution for the whole country.
Therefore, we take the vote-shares of the Clinton campaign group for each county and calculate an individual binomial distribution.
We then extract numbers of votes for Trump and Clinton to normalize them according to the total number of cast votes in each county.
For that, we use a data set from ``Dave Leip's Election Data'' \cite{DaveLeip}.

\begin{figure}
\centering
	\begin{minipage}[t]{.49\textwidth}
		\includegraphics[scale=.9]{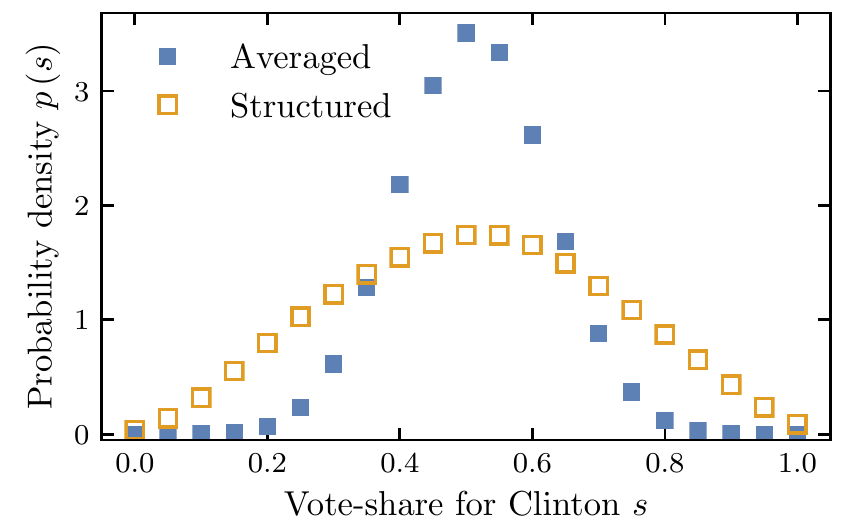}
		\label{fig:countyData_PDF}
	\end{minipage}
	\hfill
	\begin{minipage}[t]{.49\textwidth}
		\includegraphics[scale=.9]{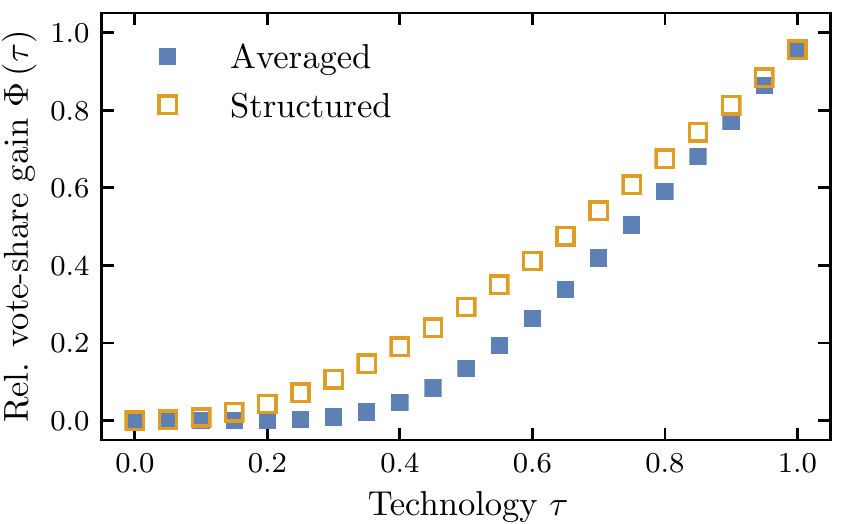}
		\label{fig:countyData_expVSG}
	\end{minipage}
	\caption{
		\textbf{Binomial distribution in comparison to actual county data.}
		The left panel shows the probability distribution of the vote-share for Clinton voters in the U.S.~presidential election based on the corresponding county data if one activist first chooses a county and then tries to persuade $k+1$ voters.
		The binomial distribution of individuals that underlies our model with the same degree only qualitatively captures the actually observed voter distribution.
		We assumed a network of degree $k+1 = 20$.
		Both distributions have the same mean, but different standard deviations.
		The right panel shows the relative vote-share gain $\Phi \left( \tau \right)$ depending on the technology $\tau$ for a binomial and the observed voter distribution. 
		The data is taken from Ref.~\cite{DaveLeip}
		For Clinton, we assume that her campaign group has a technology below the critical threshold $\tau_{B} \lesssim 0.489$.
		}
	\label{fig:countyData}
\end{figure}
We compare the probability density function of Clinton voters on county level with the global binomial distribution underlying our campaign model as defined by Eq.~\eqref{eq:voterFlow}.
For the model with county structure, we take the vote-shares of Trump and Clinton of every county and determine the corresponding binomial distributions of sample size $k+1$.
Finally, we average the distribution of the counties weighted relative to the cast vote count of each county.

Fig.~\ref{fig:countyData} (left) shows that the probability density function of the model with counties exhibits a larger standard deviation and a lower concentration of vote-shares around the mean compared to the global mean-field model.
The flatness of the probability density function of the county approach leads to an initially faster growing cumulative distribution function and thus, to a lower technology threshold for the expected relative vote-share gain as defined by Eq.~\eqref{eq:phi_tau}.
These results are illustrated in Fig.~\ref{fig:countyData} (right) and agree well with our intuition:
Real vote-share distributions are not characterized by an extreme concentration around the mean but are distributed more uniformly.
For that reason, we also expect the value of $\tau$ obtained in the parameter estimation for the last U.S.~presidential election to be smaller when considering the vote-share distribution of the individual counties.
By equating the expected vote-share gain across the distributions, the corresponding estimate is $\tau_{\mathrm{empirical}} \approx 0.75$, and thus 26.1\% larger than the value of Clinton's competing campaign group.
Including detailed information about the counties, the vote-share distribution in our model thus provides a lower and likely more plausible parameter estimation. 
These results can also be obtained by considering a network with different degrees and this leads to an even broader distribution~\cite{boettcher171}.

A similar exercise can be performed solely for the counties of the so-called battleground states.
These distributions hardly differ.
In App.~\ref{app:battlegroundstates}, we show a comparison of the distribution and the expected vote-share gain for the whole country and the battleground states only.
\section{Discussion} \label{sec:discussion}
Political campaigns are constantly undergoing transformations from one level of sophistication to the next.
In recent years, campaign groups relied more and more on technologies that allow them to maximize their vote-share gains by microtargeting voters.
We proposed and analyzed a new framework which accounts for advantageous targeting technologies to be used during political campaigns.

We have demonstrated that a technological advantage leads to substantial vote-share gains if the technology is effective enough to surpass a threshold. Furthermore, we have derived a relation between technological precision, campaign budgets, activist fractions and the campaign outcome. The application of our model to the last U.S.~presidential election shows that a technological advantage could have been crucial for winning.
This allows to gain important insights into campaign dynamics.
We demonstrated the possibility of including empirical vote-share distributions in our model to obtain better parameter estimations.

Up to this point, we focused on a pure activist-voter interaction to study the influence of targeting technologies in campaign dynamics.
Future studies should also consider the direct interaction of voters with each other, combined with the rules of the U.S.~Electoral College \cite{Fernandez-Garcia_2014_VoterModel} and including more than two parties \cite{Michaud_2018_Social}.

\section*{Acknowledgments}
The authors thank Jan Wey for valuable feedback on the manuscript. 
MH thanks Società Italiana di Fisica and Uppsala Univeristy, in particular Attila Szilva,  for the chance to present our work.
HJH thanks  CAPES and FUNCAP.

\pagebreak
\appendix
\renewcommand\thefigure{\thesection.\arabic{figure}}    
\setcounter{figure}{0}
\section{Relative budget dependence on technology} \label{app:caseDifferentiation}
\subsection{Unique steady state}\label{app:uniqueness}
The stationary state of Eq.~\eqref{eq:voterFlow} is given by $\dot{a}^0=0$ and leads to
\begin{equation}\label{eq:relativeBudgetappendix}
	\frac{a^+}{b^+} = \frac{\rho_B}{\rho_A}
	\frac{	
		\frac{ \sum_{j=0}^{k + 1} j \binom{k + 1}{j} {a^0_\mathrm{st}}^j {b^0_\mathrm{st}}^{k + 1-j}}{\sum_{j=0}^{k + 1} \binom{k + 1}{j} {a^0_\mathrm{st}}^j {b^0_\mathrm{st}}^{k + 1-j}}
		}{
		\frac{\sum_{j=\lceil \tau \left(k + 1\right) \rceil}^{k + 1} j \binom{k + 1}{j} {b^0_\mathrm{st}}^j {a^0_\mathrm{st}}^{k + 1-j}}{\sum_{j=\lceil \tau \left(k + 1\right) \rceil}^{k + 1} \binom{k + 1}{j} {b^0_\mathrm{st}}^j {a^0_\mathrm{st}}^{k + 1-j}}
		}
	=
	\frac{\rho_B}{\rho_A}
	\frac{
		\left(k + 1\right) a^0_{\mathrm{st}}
		}{
		\frac{\sum_{j=\lceil \tau \left(k + 1\right) \rceil}^{k + 1} j \binom{k + 1}{j} {b^0_\mathrm{st}}^j {a^0_\mathrm{st}}^{k + 1-j}}{\sum_{j=\lceil \tau \left(k + 1\right) \rceil}^{k + 1} \binom{k + 1}{j} {b^0_\mathrm{st}}^j {a^0_\mathrm{st}}^{k + 1-j}}
		},
\end{equation}
where $a^0_{\mathrm{st}}$ and $b^0_{\mathrm{st}}$ are the stationary states of persuadable individuals in state $A$ and $B$ respectively.
The second equality is a consequence of the binomial theorem: $\sum_{j=0}^{k + 1} \binom{k + 1}{j} x^j y^{k + 1-j}=(x+y)^{k + 1}$.
In the case of $y=1-x$, we find $\sum_{j=0}^{k + 1} \binom{k + 1}{j} x^j (1-x)^{k + 1-j}=1$. 
Taking the derivative with respect to $x$ yields for the steady state
\begin{equation}\notag
	(1-x)\sum_{j=0}^{k + 1} j \binom{k + 1}{j} x^{j-1} (1-x)^{k-j}-x \sum_{j=0}^{k + 1} \binom{k + 1}{j} (k + 1-j) x^{j-1} (1-x)^{k-j}
	=
	0.
\end{equation}
 This finally leads to 
\begin{equation}\notag
	\frac{\sum_{j=0}^{k + 1} j \binom{k + 1}{j} x^j (1-x)^{k + 1-j}}{\sum_{j=0}^{k + 1} \binom{k + 1}{j} x^j (1-x)^{k + 1-j}}
	=
	\left(k + 1 \right) x 
\end{equation}
and explains the second equality in Eq.~\eqref{eq:relativeBudgetStep2} by identifying $x$ with $a^0_{\mathrm{st}}$.
Since $b_\mathrm{st}^{0} = 1 - a_\mathrm{st}^{0}$, Eq. \eqref{eq:relativeBudgetappendix} uniquely determines the stationary solution.
Moreover, by using the proof presented in Ref.~\cite{Bottcher2018}, the model converges to a unique steady state for any initial value $a^{0} \left( 0 \right)$ and $b^{0} \left( 0 \right)$.

\subsection{Approximation}\label{app:approximation}
Due to the threshold $\lceil \tau \left(k + 1\right) \rceil$, we cannot express the denominator as a closed analytical expression but we apply a Gaussian integral approximation for the case where $k \gg 1$.
Thus, let $C(k + 1,j)$ be the binomial coefficient and $C(k + 1,j) x^j y^{k + 1-j} \sim \frac{1}{\sqrt{2 \pi \sigma^2}} \exp{\left[-\frac{(j-\mu)^2}{2 \sigma^2}\right]}$ for $k \gg 1$, where $\mu=\left(k + 1\right) x$ and $\sigma^2 = \left(k + 1\right) x y$ \cite{mathematical_methods}.
 
\begin{figure*}
	\centering
	\begin{minipage}{0.49\textwidth}
		\centering
		\includegraphics[scale=.9]{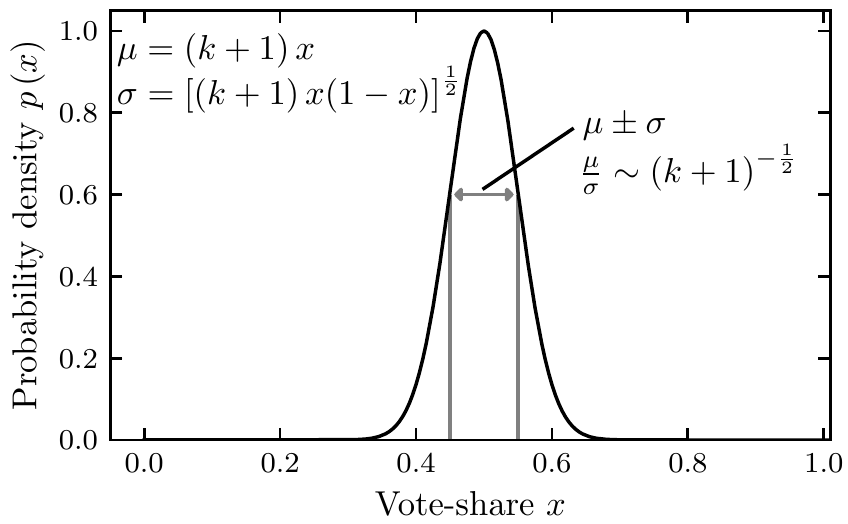}
	\end{minipage}
	\hfill
	\begin{minipage}{0.49\textwidth}
		\centering
		\includegraphics[scale=.9]{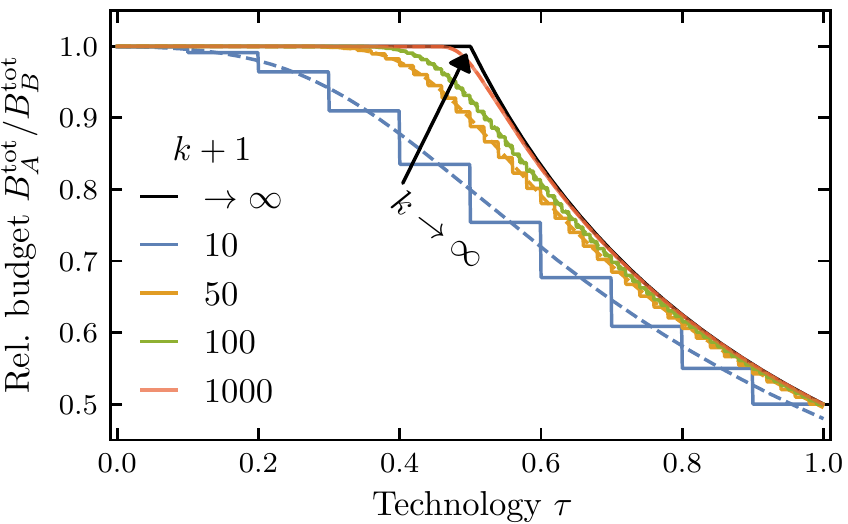}
	\end{minipage}
	  \caption{
	  	\textbf{Gaussian approximation for large degrees.}
	  	The left panel illustrates the Gaussian approximation as defined in Eq.~\eqref{eq:gaussApprox}.
	  	Since $\mu= \left(k + 1\right)x$ and $\sigma=\left[\left(k + 1\right) x (1-x)\right]^{\frac{1}{2}}$, the width of the probability density function relative to the domain is proportional to $\left(k + 1\right)^{-\frac{1}{2}}$.
	  	In the right panel, we illustrate the dependence of the activist fraction $a^+/b^+$ on $\tau$ according to Eq.~\eqref{eq:appendix_activist_tau}.
	  	The stepped solid lines and the dashed lines correspond to solutions of Eq.~\eqref{eq:relativeBudgetappendix} and Eq.~\eqref{eq:gaussApprox} for different $k + 1\in\left\{10, 50, 500, 1000\right\}$, respectively.
	  	The black arrow indicates the direction in which $k + 1$ increases towards infinity.
	  	The used parameters are $a_\mathrm{st}^{0}=b_\mathrm{st}^{0}=0.5$ and $\rho_A=\rho_B$.}
	 \label{fig:gauss_convergence}
\end{figure*}

We now set $x=b^0_\mathrm{st}$ and $y=1-b^0_\mathrm{st}$ and apply the latter approximation to the denominator of Eq.~\eqref{eq:relativeBudgetappendix} and obtain
\begin{equation}
	\frac{\int_{\tau \left(k + 1\right)}^{\infty}j ~\exp{\left[-\frac{(j-\mu)^2}{2 \sigma^2}\right]}\mathrm{d}j}{\int_{\tau \left(k + 1\right)}^{\infty} \exp{\left[-\frac{(j-\mu)^2}{2 \sigma^2}\right]}\mathrm{d}j
	}=\frac{\int_{\left(k + 1\right) \lambda}^{\infty} \left(\tilde{j}+\mu\right) ~\exp{\left(-\frac{\tilde{j}^2}{2 \sigma^2}\right)}\mathrm{d}\tilde{j}}{\int_{\left(k + 1\right) \lambda}^{\infty} ~\exp{\left(-\frac{\tilde{j}^2}{2 \sigma^2}\right)}\mathrm{d}\tilde{j}},
	\label{eq:gaussApprox}
\end{equation}
where we substituted $\tilde{j}=j-\mu$ and $\lambda=\tau - x$.
We distinguish between the cases $\tau < x$ and $\tau \geq x$.
Note that $x$ corresponds to the stationary vote-share $ b^0_\mathrm{st}$ of campaign group $B$. 
The emergence of these two cases is illustrated in Fig.~\ref{fig:gauss_convergence} (left).
We find that the width of the distribution relative to the domain decays as $\mu/\sigma \sim \left(k + 1\right)^{-\frac{1}{2}}$.
Thus, in the case of large degrees $k$, an integration defined by Eq.~\ref{eq:gaussApprox} from $\left( k + 1 \right)x<\mu$ to infinity approaches $\mu$. 

More specifically, if $\tau < x$, we have $\lambda < 0$ and the lower bound of the integrals tends towards minus infinity for large values of $k + 1$ and hence
\begin{equation}
	\frac{\int_{-\infty}^{\infty} \left(\tilde{j}+\mu\right) ~\exp{\left(-\frac{\tilde{j}^2}{2 \sigma^2}\right)}\mathrm{d}\tilde{j}}{\int_{-\infty}^{\infty} ~\exp{\left(-\frac{\tilde{j}^2}{2 \sigma^2}\right)}\mathrm{d}\tilde{j}}=\mu 
	=
	\left(k + 1\right) x.
	\label{eq:gaussApproxeval1}
\end{equation}
In this case, the relative activist fraction is $a^+/b^+=\left(\rho_B a_\mathrm{st}^0\right)/\left(\rho_A b_\mathrm{st}^0\right)$.

On the other hand, if $\tau \geq x$, both integrals of Eq.~\eqref{eq:gaussApprox} tend to zero.
For $\tau > x$, we obtain
\begin{align}
	&\lim_{k \rightarrow \infty} \frac{1}{k + 1}\frac{\int_{\left(k + 1\right) \lambda}^{\infty} \left(\tilde{j}+\mu\right) ~\exp{\left(-\frac{\tilde{j}^2}{2 \sigma^2}\right)}\mathrm{d}\tilde{j}}{\int_{\left(k + 1\right) \lambda}^{\infty} ~\exp{\left(-\frac{\tilde{j}^2}{2 \sigma^2}\right)}\mathrm{d}\tilde{j}}\\
	=&
	\lim_{k \rightarrow \infty} \frac{1}{k + 1} \left\{\sqrt{\frac{2}{\pi}}\frac{\exp\left[-\left(\left(k + 1\right) \lambda\right)^2/(2 \sigma^2)\right] \sigma}{\erfc\left[\left(\left(k + 1\right) \lambda\right)/(\sqrt{2} \sigma)\right]}+\mu\right\}\\
	=&
	\lambda+x=\tau.
	\label{eq:gaussApproxeval2}
\end{align}
In the last step, we only considered the dominant term of the following asymptotic expansion of the complementary error function $\erfc(x)$ for large arguments $x$ \cite{mathematical_methods2}:
\begin{equation*}
	\erfc \left( x \right)
	:=
	\frac{\exp\left(-x^2\right)}{\sqrt{\pi} x}\left[1+\sum_{n=1}^\infty (-1)^n \frac{1\cdot 3 \cdot 5  \cdot\dots\cdot  (2 n -1)}{(2 x^2)^n}\right].
\end{equation*}
For $\tau=x$, a similar derivation using the left-hand side of Eq.~\eqref{eq:gaussApprox} leads to the same result.
We insert the latter result into Eq.~\eqref{eq:relativeBudgetappendix} and find $a^+/b^+ = (\rho_B a_\mathrm{st}^0)/(\rho_A \tau)$ for $\tau \geq x$.
Thus, after re-substitution, the relative budget defined in Eq.~\eqref{eq:relativeBudget}, and the relative activist fraction exhibit the following dependence on the technological precision $\tau$ of campaign group $A$ for $k \gg 1$:
\begin{equation}
	\frac{B^{\mathrm{tot}}_A(T)}{B^{\mathrm{tot}}_B(T)}=
	\frac{c_A a^+}{c_B b^+} = \frac{c_A \rho_B}{c_B \rho_A} \cdot
	\begin{cases}
		\frac{a_\mathrm{st}^0}{b_\mathrm{st}^0} &\text{if } \tau < b_\mathrm{st}^0 = 1 - a_\mathrm{st}^0\\
		\frac{a_\mathrm{st}^0}{\tau} &\text{if } \tau \geq b_\mathrm{st}^0 = 1 - a_\mathrm{st}^0\\
	\end{cases}~.
	\label{eq:appendix_activist_tau}
\end{equation}

\subsection{Role of the network degree}\label{app:degree}
Finally, we briefly discuss the influence of $k + 1$ on the  goodness of fit of the Gaussian approximation used to analytically evaluate Eq.~\eqref{eq:relativeBudgetappendix} in terms of Eq.~\eqref{eq:gaussApprox}. 
More specifically, we shall focus on the following two questions: How large should the degree $k$ be (i) to approximate the binomial sum of Eq.~\eqref{eq:relativeBudgetappendix} with a Gaussian integral and (ii) to evaluate the integrals of Eq.~\eqref{eq:gaussApprox} using the approximations of Eqs.~\eqref{eq:gaussApproxeval1} and \eqref{eq:gaussApproxeval2}?

To discuss questions (i) and (ii), we now focus on Fig.~\ref{fig:gauss_convergence} (right), which illustrates the fraction $a^+/b^+$ as a function of $\tau$ for different degrees $k$.
The stepped solid lines are the solutions obtained by directly evaluating the binomial sums of Eq.~\eqref{eq:relativeBudgetappendix}, whereas the dashed lines are the solutions of the corresponding Gaussian integrals as defined in Eq.~\eqref{eq:gaussApprox}.
One sees that the Gaussian approximation describes the binomial sum well for values of $k + 1 \gtrsim 40$, which provides an answer to question (i). 
To discuss question (ii), we focus on the asymptotic solution for $k \rightarrow \infty$, which is illustrated by a solid black line in Fig.~\ref{fig:gauss_convergence} (right).
In the case of $k + 1=50$, the deviations from the asymptotic solution are substantially smaller compared to the case where $k + 1=10$. If $k + 1\gtrsim 1000$, the deviations from the asymptotic solution are negligible.
\section{Necessary technology in the battleground states to win the election}
\label{app:battlegroundstates}
Sec.~\ref{sec:crosschecking} shows that a more detailed analysis representing each county structure of the United States of America by its own mean field model allows us to deduce a lower value for the technology $\tau$ compared to a single model representing the whole country.
We now show that we obtain the same result if we only consider the data of the battle ground states (i.e., the states in which the major part of the campaigns takes place).

\begin{figure}
	\centering
	\begin{minipage}[t]{.49\textwidth}
		\includegraphics[scale=.9]{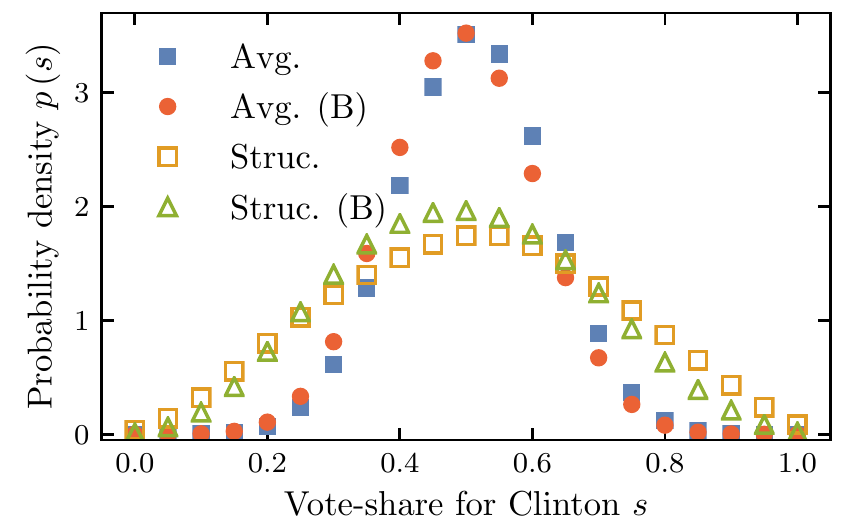}
	\end{minipage}
	\hfill
	\begin{minipage}[t]{.49\textwidth}
		\includegraphics[scale=.9]{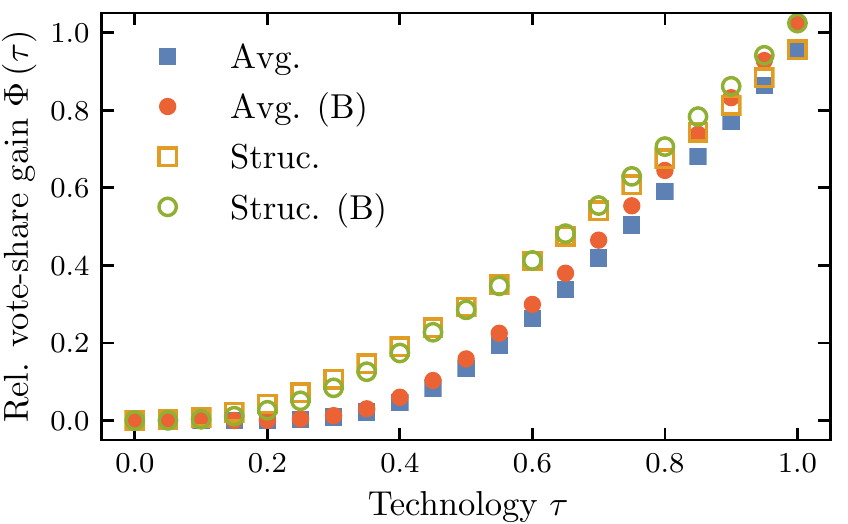}
	\end{minipage}
	\caption{
		\textbf{Binomial distribution in comparison to actual county data.}
		The left panel shows the probability distribution of the vote-share for Clinton voters in the U.S.~presidential election first for the mean-field model and second for a structured combination of mean-field model for each county based on the corresponding county data if one activist first chooses a county and then tries to persuade $k+1$ voters.
		Square shaped markers show the solution for all states and circle-shaped markers show only the battleground states.
		Filled markers show the mean-field solution (label ``Avg.'') and empty markers show the solution of the structured model (labeled ``Struc.'').
		The binomial distribution of individuals that underlies our model with the same degree 
		only qualitatively captures the actually observed voter distribution.
		We assumed a network of degree $k+1 = 20$.
		Both distributions have the same mean, but different standard deviations.
		The right panel shows the relative vote-share gain $\Phi \left( \tau \right)$ depending on the technology $\tau$ for a binomial and the observed voter distribution. 
		The data is taken from Ref.~\cite{DaveLeip}.
	}
	\label{fig:countyDataapp}
\end{figure}
The average vote-share of the popular vote for Clinton is $51.1\%$ over the whole country and $49.4\%$ in the battleground states.
We calculate the vote-share distribution for Clinton as described in Sec.~\ref{sec:crosschecking}.
We also assume the same network degree of $k + 1 = 20$ as in Fig.~\ref{fig:countyData}.
Fig.~\ref{fig:countyDataapp} (left) shows that the distribution of the vote-shares in the battleground states is slightly more skewed to the left than the single mean-field model as the mean is $1.7\%$ lower.
The same is valid for the distributions of the individual mean-field models for each county.
Therefore, the spread of the distribution of model representing the county structure is greater than the one of the single network.
Fig.~\ref{fig:countyDataapp} (right) shows that the subsequent relative vote-share gain starts to increase at significantly smaller values of the technology in model including the county structure  than in the single mean-field models.
From that taking the sub-structure into account, we conclude that a $5\%$ smaller technology is sufficient to achieve the election outcome than in a single mean-field model considering either the whole country or the battleground states only.
\end{document}